\documentclass[aps,showpacs,graphicx]{revtex4} 

\usepackage{amssymb} 
\usepackage[dvips]{graphicx} 
\usepackage[english]{babel} 
\usepackage{braket}
\newcommand{\beq}{\begin{equation}} 
\newcommand{\eeq}{\end{equation}} 
\newcommand{\beqn}{\begin{eqnarray}} 
\newcommand{\eeqn}{\end{eqnarray}} 
\newcommand{\bsigma}{\mbox{\boldmath $\sigma$}} 
\newcommand{\btau}{\mbox{\boldmath $\tau$}} 
\newcommand{\half}{\frac{1}{2}} 

\newcommand{\br}{{\bf r}}

\newcommand{\ripm}{\rho^{\rm IPM}}
\newcommand{\rsrc}{\rho^{\rm SRC}}
\newcommand{\rlrc}{\rho^{\rm LRC}}
\usepackage[usenames]{color} 
\usepackage{ulem}

\begin{document} 
 
\noindent 
\title{Effect of short- and long-range correlations on \\neutron
skins of various neutron-rich doubly magic nuclei} 
 
\author{G. Co'$^{\,1,2}$, M. Anguiano$^{\,3}$, A. M. Lallena$^{\,3}$ }
\affiliation{$^1$ Dipartimento di Matematica e Fisica ``E. De Giorgi'', 
  Universit\`a del Salento, I-73100 Lecce, ITALY, \\ 
$^2$ INFN Sezione di Lecce, Via Arnesano, I-73100 Lecce, ITALY, \\ 
$^3$ Departamento de F\'\i sica At\'omica, Molecular y 
  Nuclear, Universidad de Granada, E-18071 Granada, SPAIN
}  

\date{\today} 

\bigskip 
 
\begin{abstract} 
We study the effects of correlations beyond the independent particle model 
in the evaluation of neutron skins of various neutron-rich doubly magic nuclei. 
We consider short-  and long-range correlations to take into account the presence 
of the strongly repulsive core of the bare nucleon-nucleon interaction 
and collective nuclear phenomena, respectively. 
Despite the strong sensitivity on the structure of the nucleus considered,  
our results indicate that, in general, correlations increase the values of the neutron skins.
\end{abstract}

\bigskip 
\bigskip 
\bigskip 
 

\maketitle 
\section{Introduction}

The amount of information about the neutron density distribution 
in atomic nuclei is very poor in comparison to that of protons. 
Neutron densities have been investigated mainly with 
hadronic probes, protons \cite{sta94,cla03,zen10}, neutrons \cite{dev81,hij94}, 
$\alpha$ particles \cite{gil76,kra04}, and pions \cite{pre81,tar14},  
and the interpretation of the observed data is often affected by strong dependencies on the model considered. In contrast,
the interaction between electromagnetic probes and the nucleus is much better controlled. 
This fact triggered the idea of using polarized electron beams to study neutron skins, i.~e.
the difference between the root mean square (rms) radii of the neutron and proton density distributions \cite{thi19}.  

The first experiment of this kind was performed in the Hall A of the Thomas Jefferson
National Accelerator Facility (JLab) and investigated the $^{208}$Pb nucleus \cite{alc04}. 
The first results of this experiment were published in 2012 \cite{abr12} and 
they are indicated as PREX-1. A second campaign of data taking, 
called PREX-2, was carried out in 2021 \cite{adh21}. The combined analysis of the two 
PREX experiments provides a value of the $^{208}$Pb neutron skin of 
\beq
R_{\rm skin}(^{208}{\rm Pb}) \, = \, R_\nu(^{208}{\rm Pb}) \, - \, R_\pi(^{208}{\rm Pb}) \, = \, (0.283 \pm 0.071)\,{\rm fm} \, ,
\label{eq:Pb-skin}
\eeq
where $R_\nu$ and $R_\pi$ indicate the neutron and proton rms radii, respectively. 
This value is compatible with that of Ref.~\cite{gil76},  measured in
$\alpha$ scattering. 
In contrast, it is remarkably larger than those obtained from the
scattering with other  hadronic probes \cite{kar02,klo07,zen10,kra13,tar14},  the study of 
the electric dipole polarizability of neutron rich nuclei \cite{pie21},
the pigmy dipole resonances \cite{kli07,car10},  
the exotic atoms spectroscopy \cite{trz01,klo07,fri09,mak20}, 
the astrophysical constraints \cite{ess21},  
and those found in the great majority of the mean-field calculations 
\cite{fur02,cen10,meu14}.

\section{The mean-field model}

In this work, we analyze the reliability of  the theoretical predictions.
As it is clearly pointed out in Ref.~\cite{pie21}, these predictions have 
been done mainly in the framework of the mean-field, or independent particle,
model (IPM).
This is relevant because, within this model, the neutron skin has been strongly correlated to an important quantity describing
the nuclear matter equation of state: the slope of the density dependence of the symmetry energy at the saturation point, usually called $L$ \cite{bro00,vid09,cen09,roc13,pie21}. 
The value of $L$ has important consequences on our understanding of the structure of neutron stars \cite{fat12}. 

In our study, we calculated the proton and neutron density distributions, defined as
\beq
\rho_\alpha (\br)\, =\, \frac {A}{\langle \Psi | \Psi \rangle} \, \Bigl\langle \Psi \Bigl| \displaystyle {\sum_j}^\prime \, \delta(\br-\br_j) \Bigr| \Psi \Bigr\rangle \, , \,\,\, \alpha \, \equiv \, {\rm \pi, \nu} \, ,
\label{eq:rho}
\eeq
by considering different ans\"atze for $\ket{\Psi}$, the wave function describing the nuclear ground state. In Eq.~(\ref{eq:rho}),
the prime indicates that the sum is restricted to protons or neutrons only, according to $\alpha$. 
The rms radii $R_\nu$ and $R_\pi$  and, consequently, the neutron skins, were obtained
by using $\rho_\nu$ and $\rho_\pi$ as
\beq
R_\alpha\, \, = \, \left[ \displaystyle \frac 
{\displaystyle \int_0^\infty {\rm d}r \, r^4 \, \rho_\alpha (r)}
{\displaystyle \int_0^\infty {\rm d}r \, r^2 \, \rho_\alpha (r)} 
\right]^{\half} \, ,  \,\,\, \alpha \, \equiv \, \pi, \nu \, .
\label{eq:radius}
\eeq
We found that effects beyond the IPM, the correlations, 
affect differently the neutron and proton density distributions and, therefore, 
modify the IPM neutron skin values. 

In the IPM, $\ket{\Psi}$ is a Slater determinant of single-particle (s.p.) wave functions.
In this model, each nucleon moves independently of the other ones, the only limitation 
being that imposed by the Pauli exclusion principle. 
We constructed this IPM collective state by solving a set of Hartree-Fock (HF) equations with
a  density dependent effective finite-range nucleon-nucleon interaction. 
Specifically, we considered two parametrizations of the Gogny interaction, the so-called
 D1S  \cite{ber91} and D1M \cite{gor09} forces. We also carried out calculations with an interaction 
 containing tensor terms, the D1ST2a \cite{ang12}, but, since  the effects of these terms on the 
radii were within the numerical accuracy of our calculations, we do not examine here
the results obtained with this force. 

\section{Results}

In addition to the $^{208}$Pb nucleus, which is the main subject of our investigation, we considered
other neutron-rich doubly magic nuclei, $^{48}$Ca, $^{68}$Ni, $^{90}$Zr and $^{132}$Sn, to verify 
that our findings are not strictly related to some specific feature of $^{208}$Pb. 
We investigated the importance of pairing effects in these nuclei by carrying out Bardeen-Cooper-Schrieffer calculations 
with the same effective nucleon-nucleon interactions used in HF, and we found them irrelevant.

The quality of our IPM in describing the ground state of the nuclei considered is summarized in 
Table~\ref{tab:be}, where the values of binding energies and charge rms radii are compared to
the experimental values taken from the compilations of Refs. \cite{bnlw, ang13}.  
The agreement with the experimental data is not a surprise since
the values of the parameters of the two forces used in the HF calculations
were selected by doing a fit of about 2000  binding energies and 900 charge rms radii  \cite{ber91,gor09,cha07t},
and the nuclei we considered are among those chosen for the fit. 

\begin{table}[!h]
\begin{center}
\begin{tabular}{r c ccc c ccc}
\hline\hline
&~~& \multicolumn{3}{c}{$B/A$ (MeV)} &~~& \multicolumn{3}{c}{$R_{\rm ch}$ (fm)} \\ 
\cline{3-5} \cline{7-9}
&& D1S & D1M & exp. & & D1S & D1M & exp. \\ \hline
$^{48}$Ca && 8.691 & 8.590 & 8.666 & & 3.539 & 3.514 & 3.477  \\
$^{68}$Ni && 8.648 & 8.584 & 8.682 & & 3.923 & 3.894 & -  \\
$^{90}$Zr && 8.739 & 8.635 & 8.709 & & 4.292 & 4.264 & 4.269 \\
$^{132}$Sn && 8.513 & 8.308 & 8.354 & & 4.672 & 4.700 & 4.709 \\ 
$^{208}$Pb && 7.895 & 7.829 & 7.867 & & 5.489 & 5.501 & 5.501 \\
\hline\hline
\end{tabular}
\caption{\small Binding energies per nucleon, $B/A$, and rms charge radii, $R_{\rm ch}$,
obtained in the IPM calculations with the D1S and D1M interactions, 
compared to the experimental data taken from the compilations of Refs. \cite{bnlw, ang13}.}
\label{tab:be}
\end{center}
\end{table}

We show in Table~\ref{tab:skin} the neutron skins obtained in the calculations carried out within our IPM. 
The values found with the D1M interaction are smaller than those calculated with the D1S force. 
The relative differences between these two types of calculations 
are $24.4\%$ for $^{208}$Pb, 
about $15\%$, for $^{68}$Ni, 
$^{90}$Zr and $^{132}$Sn, and $7\%$ for $^{48}$Ca. 

\begin{table}[!t]
\begin{center}
\resizebox{\columnwidth}{!}{
\begin{tabular}{r c c c cccc c cccc}
\hline\hline
&~~&&~~& \multicolumn{4}{c}{D1S} &~~& \multicolumn{4}{c}{D1M} \\ 
\cline{5-8} \cline{10-13}
&& $(N-Z)/A$ && IPM & LRC & SRC & total && IPM & LRC & SRC & total \\ 
\hline
$^{ 48}$Ca &&   0.167 && 0.145 & 0.175 ($ 20.2\%$) & 0.157 ($  8.2\%$) & 0.186 ($ 28.3\%$) && 0.134 & 0.162 ($ 20.3\%$) & 0.147 ($  9.2\%$) & 0.174 ($ 29.5\%$) \\ 
$^{ 68}$Ni &&   0.176 && 0.157 & 0.224 ($ 42.1\%$) & 0.182 ($ 16.0\%$) & 0.248 ($ 57.6\%$) && 0.135 & 0.166 ($ 23.0\%$) & 0.145 ($  7.3\%$) & 0.176 ($ 30.2\%$) \\ 
$^{ 90}$Zr &&   0.111 && 0.058 & 0.055 ($ -5.4\%$) & 0.062 ($  7.8\%$) & 0.059 ($  1.9\%$) && 0.050 & 0.034 ($-32.1\%$) & 0.055 ($  9.8\%$) & 0.039 ($-23.1\%$) \\ 
$^{132}$Sn &&   0.242 && 0.190 & 0.198 ($  4.3\%$) & 0.202 ($  6.2\%$) & 0.210 ($ 10.5\%$) && 0.163 & 0.166 ($  1.7\%$) & 0.189 ($ 15.9\%$) & 0.192 ($ 17.6\%$) \\ 
$^{208}$Pb &&   0.212 && 0.122 & 0.142 ($ 16.9\%$) & 0.144 ($ 18.4\%$) & 0.165 ($ 35.2\%$) && 0.092 & 0.109 ($ 18.8\%$) & 0.115 ($ 24.6\%$) & 0.132 ($ 43.2\%$) \\ 
\hline \hline
\end{tabular}}
\caption{\small Neutron skins, in fm, obtained in IPM and by including correlations. The calculations 
were done by using the D1S and D1M forces.  
The relative differences with respect to the IPM values are shown between parentheses.
 }
\label{tab:skin}
\end{center}
\end{table}

The almost free motion of the nucleons inside the nucleus is modified by the presence of 
effects that may be classified as of short- and long-range. The short-range correlations (SRC) 
are due to the presence of the strongly repulsive core of the bare nucleon-nucleon 
interaction and prohibit two nucleons from getting too close to each other.
The long-range correlations (LRC) take into account the part of the interaction
neglected in the HF approach, which is usually called residual interaction
and couples collective vibrations to the s.p. wave functions \cite{row70,rin80,suh07}.

\begin{figure}[!b]
\hspace*{0.2cm}
\begin{minipage}[c]{0.45\linewidth}
\includegraphics[width=4.5cm,angle=90]{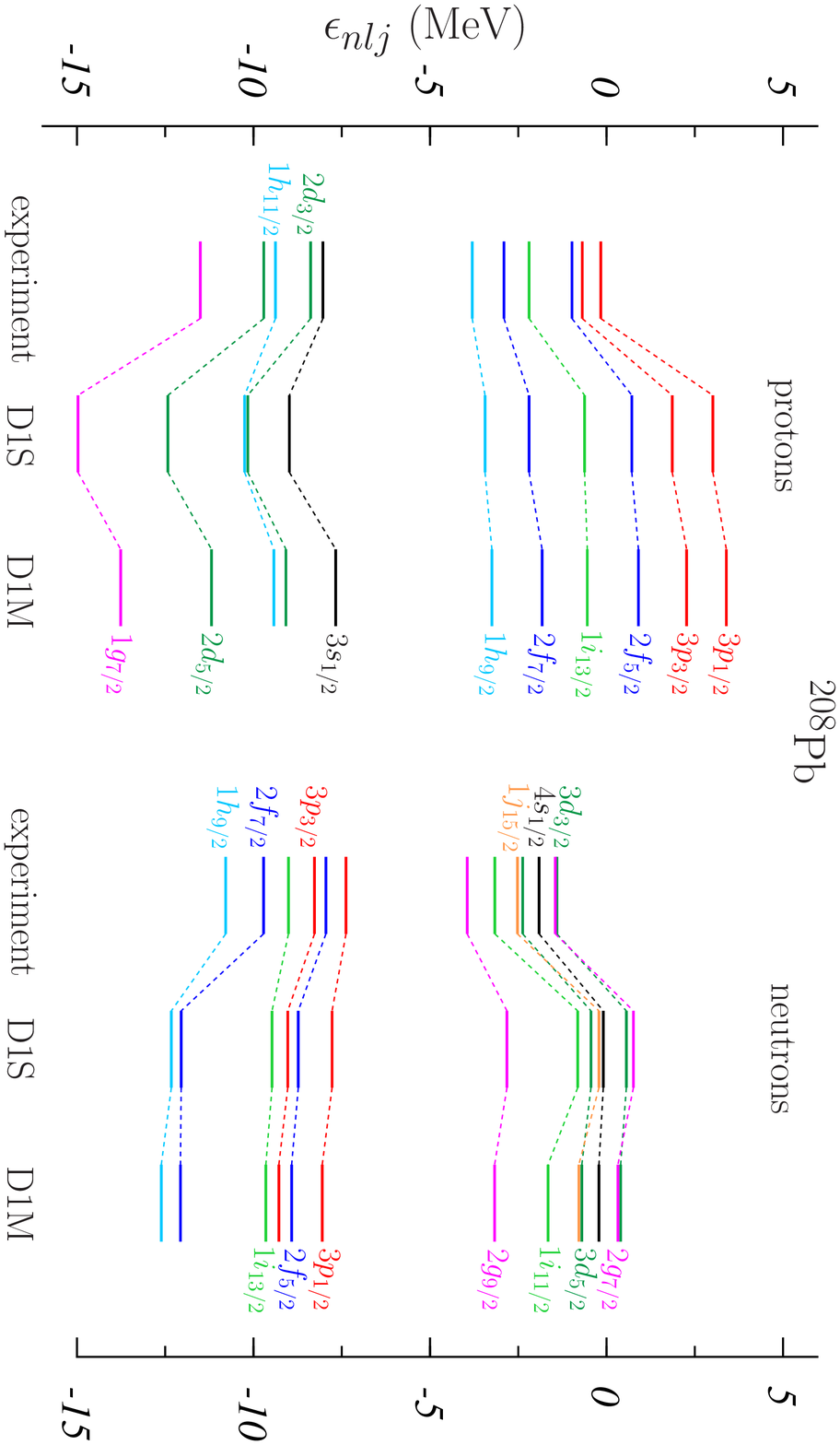} 
\vspace{-0.3cm}
\caption{\small Energies, in MeV, of the s.p. levels of $^{208}$Pb close to the Fermi surface.
}
\label{fig:spene} 
\end{minipage}
\hspace*{0.5cm}
\begin{minipage}[c]{0.45\linewidth}
\vspace*{0.75cm} %\hspace*{-8cm}
%\hspace*{-0.2cm}
\includegraphics[width=6.5cm,angle=0]{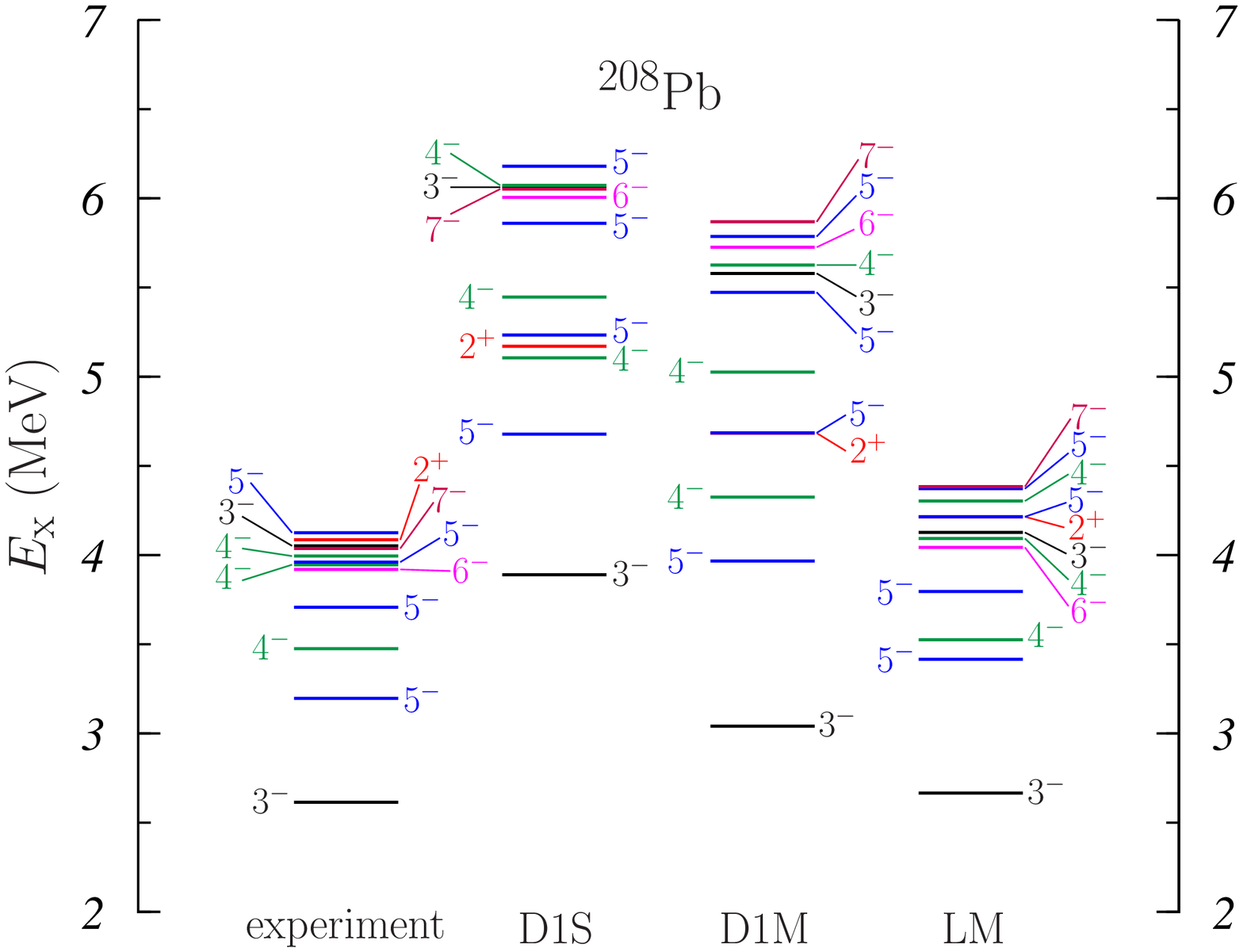} 
\vspace{-0.3cm}
%\begin{center}
\caption{\small 
Excitation spectrum of $^{208}$Pb obtained by self-consistent RPA calculations 
carried out with D1S and D1M interactions 
and by the phenomenological LM approach.  
}
\label{fig:spec} 
%\end{center} 
\end{minipage}
\end{figure}

\subsection{Long-range correlations}

We treated LRC within the theoretical framework of the random phase approximation (RPA)
\cite{row70,rin80,suh07}. In this theory, the nuclear ground state is no longer the IPM Slater determinant 
but a more complicated state containing a set of 
particle-hole excitations that are weighted by the so-called backward amplitudes, $Y$, 
obtained by solving the RPA equations. In Refs. \cite{len90,ang01,co22} this idea has been exploited
to evaluate nuclear ground state properties. The correlated density distributions can be expressed as \\
\beq
\nonumber
\rlrc_\alpha(r) \, = \, \ripm_\alpha(r) \,- \, \sum_{J^\Pi} \, \frac{2J+1}{8\pi} \, \sum_{E_k} \, 
{\sum_{\rm p,h}}^\prime \, \left| Y_{\rm p h}^{J^\Pi} (E_k) \right|^2  \left\{
\left[ {\cal R}_{\rm p}(r) \right]^2 \, - \, \left[ {\cal R}_{\rm h}(r) \right]^2  \right\} \, , 
\,\,\, \alpha \, \equiv \, \pi, \nu \, ,
\label{eq:rholrc}
\eeq
where $\ripm_\alpha$ indicates the IPM density distribution, ${\cal R}$ the radial part of the 
particle, p, or hole, h, s.p. wave function and $J$ and $\Pi$ the angular momentum 
and the parity of a specific nuclear excitation with energy $E_k$.

Our RPA calculations were carried out by consistently using the same interaction 
adopted in the HF calculations. The numerical stability of the RPA results was ensured
by following the prescriptions described in Ref.~\cite{don09}. 
For each nucleus considered we included all the multipolarities 
whose experimental excitation energy is smaller than $5\,$MeV. 
We verified that the inclusion of other multipolarities 
did not modify significantly the final result. 

For the nucleus $^{208}$Pb we also performed calculations
with the phenomenological approach of Ref.~\cite{spe77}, inspired 
to the Landau-Migdal (LM) theory of the finite Fermi systems.
In this case, the s.p. wave functions were generated by two 
Woods-Saxon potentials, one for the protons and the other one for the neutrons, 
whose parameters, given in Ref. \cite{rin78},
were selected to reproduce at best the empirical values of the s.p. energies. 
The phenomenological RPA calculations were carried out by considering these experimental values,
which are compared in Fig.~\ref{fig:spene} to the HF s.p. energies. The figure clearly
shows the well known fact that the empirical s.p. spectrum is more compressed than
that predicted by HF calculations \cite{rin80}. In this phenomenological RPA calculation
the residual interaction is a zero-range density dependent LM force whose
parameters values were selected to reproduce the excitation energy of the low-lying $3^-$ 
excitation and the position of the centroid energy of the electric monopole excitation. 
In Fig. \ref{fig:spec} we compare the excitation spectrum of the $^{208}$Pb obtained for
the three different approaches with the experimental one. As expected, the results obtained 
with the phenomenological LM interaction reproduce better the experimental data than 
those found with the self-consistent HF plus RPA calculations. 

\begin{figure}[!h]
\hspace*{0.2cm}
\begin{minipage}[c]{0.45\linewidth}
\vspace*{-0.85cm}
\includegraphics [scale=0.3,angle=0]{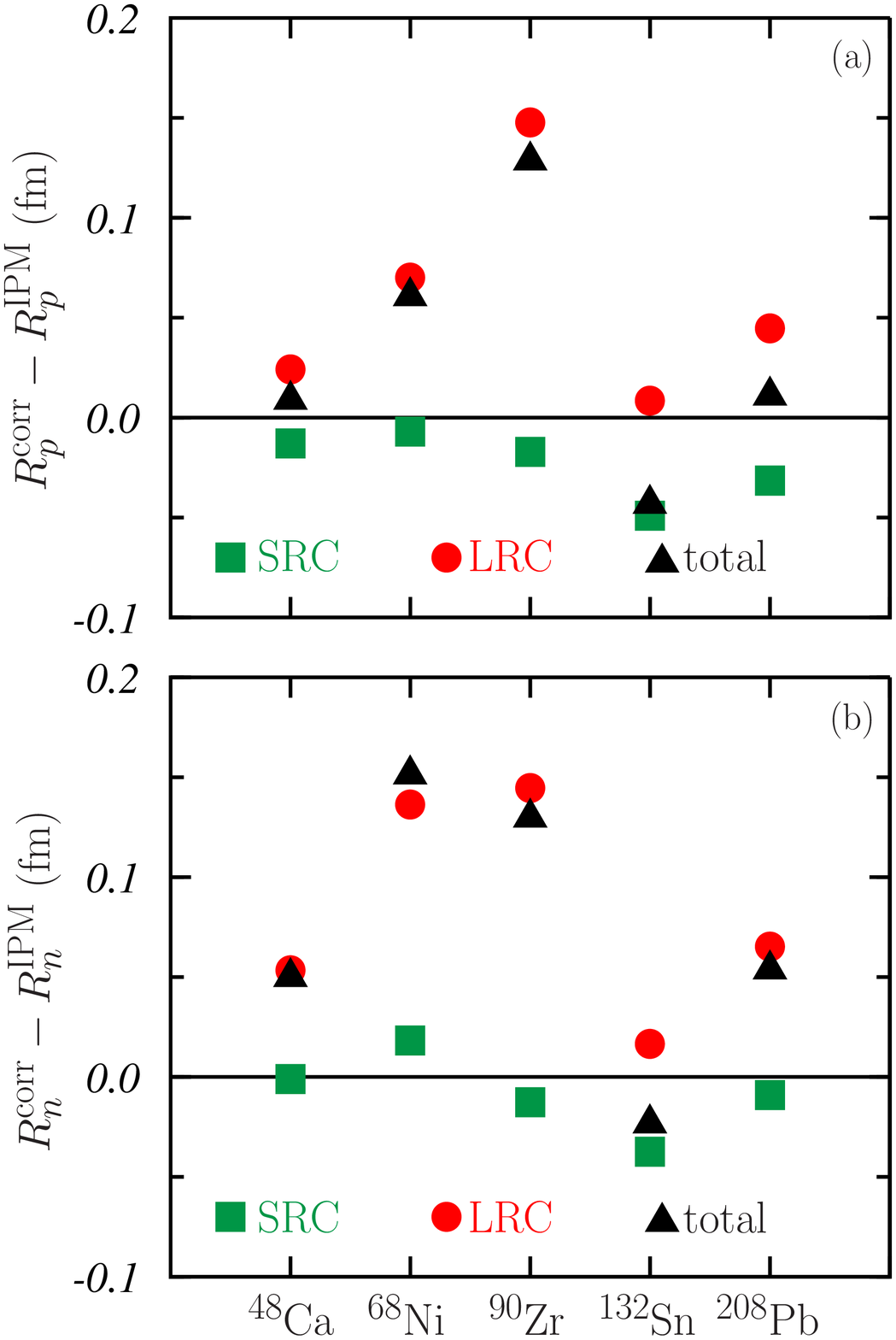} 
\vspace{-0.3cm}
\caption{\small Differences between correlated and 
IPM rms radii for (a) protons and (b) neutrons. 
All the calculations were done with the D1S interaction. 
The red circles indicate the results obtained by considering the LRC only,
the green squares those with the SRC only, and the black triangles the
total effect. 
}
\label{fig:diff} 
\end{minipage}
\hspace*{0.5cm}
\begin{minipage}[c]{0.45\linewidth}
\vspace*{0.75cm}\includegraphics [scale=0.3,angle=0]{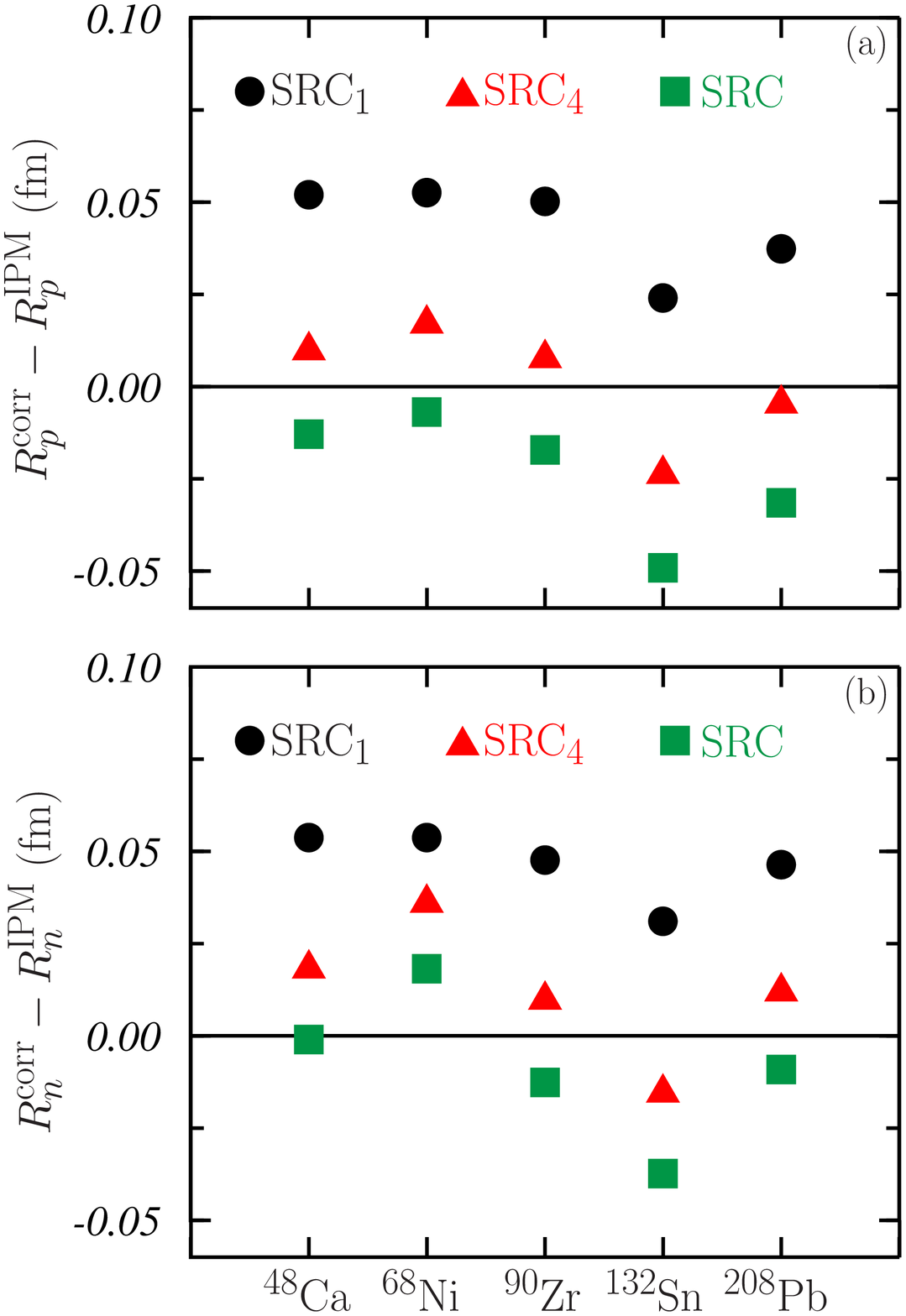} 
\vspace{-0.3cm}
\caption{\small 
Differences between the correlated rms radii and those obtained in the IPM results. 
All the calculations were done with the D1S interaction.
The green squares indicate the results obtained by considering all the six operator dependent terms of the correlation function (\ref{eq:fcorrel}), these are the same as in Fig.~\ref{fig:diff}.
The red triangles indicate the values found by considering only the first four terms of the correlation function, the so-called central terms. The black circles correspond to the results obtained when only the scalar term of the correlation function is included in the calculation. 
}
\label{fig:srcdiff} 
\end{minipage}
\end{figure}

The effects of the LRC on the proton and neutron rms radii are shown 
in Fig.~\ref{fig:diff}, where we present the results obtained with the D1S interaction. 
We found similar results by using the D1M force. In this figure, the red circles show
the differences $R_\alpha^{\rm LRC}-R_\alpha^{\rm IPM}$ for both protons 
(Fig.~\ref{fig:diff}a) and neutrons (Fig.~\ref{fig:diff}b).
In all the nuclei analyzed these differences are positive, indicating that the global effect 
of the LRC is a broadening of the IPM density distributions with the subsequent increase of the proton and neutron rms radii.

The size of this effect is not the same on proton and neutron densities.
This difference is large enough to change the value of the neutron skin. 
The IPM and LRC columns of Table~\ref{tab:skin} show that
the LRC skin values are larger than the IPM ones in all the cases, 
except for $^{90}$Zr nucleus.
In this latter case, the  LRC reduce the neutron skin of $5.4\%$, 
in the calculation with the D1S force, and of $32.1\%$, in those with the D1M. 
In the other nuclei, the relative difference ranges from $4.3\%$ for $^{132}$Sn  
to about $20\%$ for $^{48}$Ca and $^{208}$Pb. 

In the phenomenological LM calculation for $^{208}$Pb 
the LRC produce an increase of proton and neutron rms radii of about 0.02 fm,
and a relative increase of the skin of about 3\%.

The relevance of the various excitation multipoles included in the sum of Eq. (\ref{eq:rholrc}) was also evaluated. 
As expected, we found that the most important ones are those with the lowest excitation energy showing a rather collective behavior. 
This is the case of the $3^-$ excitation in $^{208}$Pb. The inclusion of this multipole
in the sum of Eq. (\ref{eq:rholrc}) modifies the proton rms radii by 0.3\% in the calculation 
with the D1S force, by 0.8\% with the D1M, and by 1.2\% with the LM interaction. 
By looking at the excitation spectrum of Fig.~\ref{fig:spec}, it becomes clear 
that the effect is larger the lower is the excitation energy of the multipole.

\subsection{Short-range correlations}

We conducted the study of the LRC in a consistent picture where the only physics input
is the effective nucleon-nucleon interaction. Our treatment of the SRC, based on 
the approach proposed in Ref. \cite{co95} and used in Refs. \cite{ari97, ang01, co22}
to study charge density distributions, required a new physics input, the two-body correlation 
function. We described the nuclear ground state as
\beq
|\Psi \rangle \, \equiv \, \Psi^{\rm SRC} (1,2,\ldots,A)\,=\, F(1,2,\ldots,A) \, \Phi (1,2,\ldots,A) \, ,
\label{eq:psisrc}
\eeq
where we indicate with $\Phi$ the HF Slater determinant, and with 
$F$ a many-body correlation function 
defined as \cite{pan79,ari07}
\beq
F(1,2,\ldots,A)\, = \, {\cal S} \, {\prod_{i<j}} \, \sum_{p=1}^6 \, f^{(p)} (r_{ij}) \, {O_{i,j}^{(p)}} \, .
\label{eq:fcorrel}
\eeq 
In the above expression ${\cal S}$ is a symmetrization operator, 
$f^{(p)}(r_{ij})$ a scalar two-body correlation function acting on the $(i,j)$
nucleon pair, and $\{O^{(p)},\,p=1,\ldots,6\}$ indicate
two-body operators classified as in the usual Urbana-Argonne sequence \cite{pan79}: 
\beqn
\nonumber
O_{ij}^{(1)} \, &=& \, 1 \, , \,\,\, 
O_{ij}^{(2)} \, = \btau(i) \cdot \btau(j) \, , \,\,\, \\
\nonumber
O_{ij}^{(3)} \, &=&\bsigma(i) \cdot \bsigma(j) \, , \,\,\,
O_{ij}^{(4)} \, =\bsigma(i) \cdot \bsigma(j) \,  \btau(i) \cdot \btau(j) \, , \\
O_{ij}^{(5)} \, &=& S(i,j) \, , \,\,\,
O_{ij}^{(6)} \,=S(i,j)\, \btau(i) \cdot \btau(j)  \, ,
\eeqn
where $\bsigma$  and $\btau$ are the spin and the isospin operators, respectively,
and $S(i,j)$ is the usual tensor operator. 

The use of the expression (\ref{eq:psisrc}) in the definition (\ref{eq:rho}) of the density distribution
allows an expansion in clusters, each of them identified by the number of two-body correlation functions
\beq
h^{(p)}(r_{ij}) \, = \, f^{(p)}(r_{ij}) \,-\, \delta_{p,1}  
\label{eq:hdef}
\eeq
that it includes. In the previous equation $\delta$ indicates the Kronecker symbol. The key point of the model of Ref.~\cite{co95} consisted in retaining only those terms of the expansion that contain a single correlation function $h^{(p)}$. Explicit expressions of the contribution of these diagrams in terms of the radial s.p. wave functions are given in Ref.~\cite{co95}. While this truncation of the cluster expansion is a very poor approximation in the evaluation of the ground state energy, it is rather good for the density distribution. Its validity was tested by comparing the results of this model with the density distributions obtained in Fermi-Hypernetted-Chain (FHNC) calculations \cite{ari07} in which
almost all the cluster terms of the expansion are considered.

Even though the D1S and D1M interactions do not contain tensor dependent terms, we 
used all the six operator components of the correlation (\ref{eq:fcorrel}). 
In order to disentangle the effects of the various terms of the correlation function, 
we carried out calculations by including only the scalar part, i.e. $f^{(1)}$, 
the first four terms, and the complete two-body correlation function. 
We call SRC$_1$, SRC$_4$ and SRC, respectively, the results obtained in these three types of calculations.

We used the two-body correlation functions $f^{(p)}$  
obtained in Ref. \cite{ari07} with a minimization procedure
that generates specific correlations for each nucleus investigated. 
In the present study, we considered the 
two-body correlation functions obtained for the $^{48}$Ca 
and $^{208}$Pb nuclei by using
the microscopic Argonne V8' two-body force plus the 
Urbana IX interaction (see Fig.~21 of Ref.~\cite{ari07}). 
These two correlation functions are very similar and produce 
results which differ by few parts on a thousand, 
therefore we show here only those obtained with the $^{208}$Pb correlation.

The effects of the SRC on the proton and neutron rms radii 
calculated with the D1S interaction can be seen in Fig.~\ref{fig:srcdiff}
where we show the differences between correlated and IPM proton and
neutron rms radii.
The SRC$_1$ results are shown by the black circles. 
We observe that the values of all the rms radii increase with respect to the IPM ones. 
The inclusion of the other operator dependent central terms of the correlation 
(see SRC$_4$ results), reduces the effect of the scalar term as it is shown 
by the red triangles in the figure. Also the remaining 
two terms of the correlation, the tensor ones, reduce the effect of the scalar correlation, 
producing the SRC shown by  the green squares. These are the same green 
squares of Fig.~\ref{fig:diff} where they are compared to the LRC results. 

It is remarkable the difference between the behaviors of LRC and SRC.
The effects of the former ones on the rms radii strongly depend on 
the structure of the nucleus considered. For example, the size of these 
effects on $^{90}$Zr is much larger than in $^{132}$Sn. 
On the contrary, the SRC effects are almost constant in all the nuclei considered, indicating that the 
short-range features are really almost independent of
the presence of the surface and shells effects.

As we have already pointed out for the LRC, 
also in the case of the SRC the effects are slightly different for protons and 
neutrons rms radii and, consequently,  the final result is an increase of the IPM neutron skin 
in the all five nuclei analyzed, as indicated by the corresponding column of Table~\ref{tab:skin}. 
The size of the effect of the SRC is of the same order of magnitude of that 
found for the LRC, even though it depends strongly on the specific features of nucleus considered. 
We obtain a minimum increase of 6.2\% for the $^{132}$Sn nucleus calculated
with the D1S force and a maximum value of 24.6\% in $^{208}$Pb with the D1M interaction.

Since the treatments of SRC and LRC are based on different grounds,
we defined the totally correlated density as
\beq
\rho_\alpha^{\rm tot} (r) \,=\, \rlrc_\alpha(r) + \rsrc_\alpha(r) - \ripm_\alpha(r) \, , \,\,\, \alpha \, \equiv \, \pi, \nu \, .
\label{eq:rhotot}
\eeq
By considering these densities we calculated the totally correlated rms radii. The differences with the IPM radii are indicated in Fig. \ref{fig:diff} by the black squares which are roughly the sum of the LRC and SRC results. The total effects of the correlations on the neutron skins 
are presented in Table~\ref{tab:skin}: it is evident that they 
produce an increase of the skin values obtained in the IPM. 
The only exception is the case of the $^{90}$Zr when the D1M force is used.

The results of our study given in Table~\ref{tab:skin} show some regularity, but 
the specific structure of each isotope is more important than any general trend.
As example of this, we observe that  size of the neutron skin increases when the value 
of the $(N-Z)/A$ ratio increases, i.e. when the neutrons become more important. 
The results of the $^{208}$Pb nucleus are out of this trend.

\section{Conclusions}

To summarize our results we can state that, in general, correlations increase the values of the
neutron skins obtained in the IPM. Our treatment of correlations distinguishes 
between long- and short-range correlations. 
While we treated the former ones with an approach completely consistent with the IPM, 
the SRC were considered by inserting an additional physics 
ingredient not constrained by the choice of the IPM. 
The presence of operator dependent terms in the correlation function reduces the
global effect of SRC with respect to results obtained with purely scalar functions.
It tuns out that the effects of the SRC in the rms radii are relatively small as compared to those 
of the LRC and they are almost identical in each nucleus we studied. 
The dominant effects of the LRC are strongly related to the structure of the nucleus considered. 
 
The correlation effects are, on average, two times larger than those related to the use of 
different effective interaction. In our calculations the values of the neutron skins obtained 
with the D1S force are always larger than those obtained with the D1M interaction. 
 
In case of $^{208}$Pb, the inclusion of correlations improves the agreement with the PREX results. The IPM
skin values are located at $2.3 \, \sigma$ and $2.7\,\sigma$ from the mean value
given by the experiment for the D1S and D1M interactions, respectively. The corresponding correlated
values are at $1.6 \, \sigma$ and $2.1\,\sigma$.

The proper manner of tackling the description of neutron skins  
is a fully consistent, and microscopic, calculation of  the density distributions
such as those carried out with the coupled cluster model for the $^{48}$Ca in Ref. \cite{hag16}.
Up to now this kind of calculations is not feasible for heavier nuclei, 
certainty not for $^{208}$Pb. To the best of our knowledge, 
the only microscopic approach that has studied $^{208}$Pb is the FHNC calculation of Ref.~\cite{ari07}. 
Unfortunately, the numerical precision of these calculations does not allow a 
sufficient accuracy to obtain reliable results for the neutron skin. 
It is worth mentioning that some preliminary test calculations of this kind 
by using Argonne V8' two-body potential  plus three-body Urbana IX 
force show a trend in agreement with our findings \cite{pud97}.

%%%%%%%%%%%%%%%%%%%%%%%%%%%%%%%%%%%%% 
\acknowledgments  
This work has been partially supported by the Consejer\'{\i}a
de Econom\'{\i}a, Conocimiento, Empresas y Universidad,
Junta de Andaluc\'{\i}a (FQM387), the Ministerio de Ciencia e
Innovaci\'on of Spain (PID2019-104888GB-I00), and the European
Regional Development Fund (ERDF). G~C. thanks Sonia
Bacca for useful discussions.

%

%\end{landscape}
%
%
% Figures
%
%
\newpage \clearpage
%-----------------------------------------
% spene
%-----------------------------------------
\newpage \clearpage

%-----------------------------------------
% spectrum
%-----------------------------------------
\newpage \clearpage
%-----------------------------------------
% differences
%-----------------------------------------
\newpage \clearpage
%-----------------------------------------
% differences1
%-----------------------------------------
\newpage \clearpage

\end{document}